\documentclass[aps,preprint]{revtex4}%
\usepackage{amsfonts}
\usepackage{amsmath}
\usepackage{amssymb}
\usepackage{graphicx}%
\providecommand{\U}[1]{\protect\rule{.1in}{.1in}}
\begin{document}
\title{Cooperating or Fighting with Decoherence\\in the Optimal Control of Quantum Dynamics}
\author{Feng Shuang}
\author{Herschel Rabitz}
\affiliation{Department of Chemistry, Princeton University, Princeton, New Jersey 08544}

\begin{abstract}
This paper explores the use of laboratory closed-loop learning control to
either fight or cooperate with decoherence in the optimal manipulation of
quantum dynamics. Simulations of the processes are performed in a Lindblad
formulation on multilevel quantum systems strongly interacting with the
environment without spontaneous emission. When seeking a high control yield it
is possible to find fields that successfully fight with decoherence while
attaining a good quality yield. When seeking modest control yields, fields can
be found which are optimally shaped to cooperate with decoherence and thereby
drive the dynamics more efficiently. In the latter regime when the control
field and the decoherence strength are both weak, a theoretical foundation is
established to describe how they cooperate with each other. In general, the
results indicate that the population transfer objectives can be effectively
met by appropriately either fighting or cooperating with decoherence.

\end{abstract}
\date{\today}
\maketitle

\section{Introduction}

\qquad Control over quantum dynamics phenomena is the focus of many
theoretical \cite{Rice00,Rabitz0364,Shapiro03} and
experimental\cite{Walmsley0343,Brixner011} studies. Various control scenarios
exist, including optimal control\cite{Rabitz884950,Kosloff89201}, coherent
control\cite{Shapiro864103}, and STIRAP(Stimulated Raman Adiabatic Passage)
control\cite{Bergmann905363, Rice982885}. Increasing numbers of control
experiments, including on complex systems\cite{Gerber98919, Gerber9910381,
Kunde00924, Bartels00164, Brixner0157, Levis01709, Herek02533, Daniel03536},
employ closed-loop optimal control\cite{Judson921500}. Many of the control
systems explored theoretically are restricted to pure state dynamics or the
dynamics of isolated quantum systems. In practice, control field
noise\cite{JM0010841,Rabitz0263405,Ignacio049009} and
decoherence\cite{Blanchard00,Braun01} inevitably will be present in the
laboratory and the general expectation is that their involvement will be
deleterious towards achieving control. Recent studies have investigated the
effect of field noise and shown that controlled quantum dynamics can survive
even intense field noise and possibly cooperate with it under special
circumstances\cite{Shuang049270}. Decoherence of quantum dynamics in open
systems, which often represents realistic situations, is a concern for control
of atomic and molecular process, especially in condensed phases. Simulations
have shown that it's possible to use closed-loop learning control to suppress
the effect of quantum decoherence\cite{Wusheng036751}. Some investigations
have been performed on laser control of population transfer in dissipative
quantum system\cite{Rabitz018867,Rice0210810,Batista02143201,Xu046600}.
Control of decoherence and decay of quantum states in open systems has also
been explored\cite{OpticsExpress980209, Viola042357}. Decoherence was shown to
possibly be constructive in quantum dynamics\cite{Prezhdo004413,
Kendon0442315}.

This paper considers the influence of decoherence (dissipation) upon the
controlled dynamics of population transfer with the decoherence induced by
interaction with the environment but spontaneous emission is not included.
This regime arises, for example, in condensed phases where significant
environmental interactions dominate the decoherence processes. The goal is to
demonstrate that effective control of population transfer is possible in the
presence of decoherence. It is naturally found that decoherence is deleterious
to achieving control if a high yield is desired, however we also find that
good control solutions can be found that achieve satisfactory yields. If a low
yield is acceptable, then it is shown that decoherence even can be helpful and
the control field can cooperate with the decoherence. This paper will
investigate these phenomena numerically and analytically to illustrate the
issues, and this work compliments an analogous study considering the influence
of field noise\cite{Shuang049270}.

The dynamical equations and control formulation is presented in Section II.
Simulations of closed-loop management of dynamics with decoherence is given in
Sec.III. Section IV develops an analytical formulation to describe how the
control field and the decoherence cooperate with each other when they are both
weak. Finally, a brief summary of the findings is presented in Section V.

\section{The Model System}

Realistic quantum systems in the laboratory often can not be fully described
by a simple, fully certain Hamiltonian. Instead, almost all real systems are
affected by dissipation, decoherence or noise, which can not be totally
eliminated. The recent successes of closed-loop optimization
algorithms\cite{Judson921500} operating in the laboratory demonstrate the
capability of finding optimal, stable and robust solutions
automatically\cite{Gerber98919, Gerber9910381, Kunde00924, Bartels00164,
Brixner0157, Levis01709, Herek02533, Daniel03536}, even for very complex
systems. The present work aims to support the continuing experiments by
investigating the principles and rules of control in the presence of
decoherence. In keeping with this goal, a simple model system will be investigated.

The effect of dissipation on controlled quantum dynamics will be explored in
the context of population transfer in multilevel systems characterized by the
dynamical equation of the reduced density matrix $\rho_{S}$%

\begin{subequations}
\label{Liou1}%
\begin{align}
\frac{\partial\rho_{S}(t)}{\partial t}  &  =-i[H_{0}-\mu E(t),\rho
_{S}(t)]+\gamma\mathcal{F}\left\{  \rho_{S}(t)\right\}  \text{,}\label{rhos}\\
H_{0}  &  =\sum_{\upsilon}\varepsilon_{\upsilon}\left\vert \upsilon
\right\rangle \left\langle \upsilon\right\vert \text{,} \label{H0}%
\end{align}
where $\left\vert \upsilon\right\rangle $ is an eigenstate of the controlled
system Hamiltonian $H_{0}$, and $\varepsilon_{\upsilon}$ is the associated
$\upsilon$-th field-free eigen-energy. The noise-free control field $E(t)$ is
taken to have the form%

\end{subequations}
\begin{subequations}
\begin{align}
E(t)  &  =s\left(  t\right)  \sum_{l}A_{l}\cos\left(  \omega_{l}t+\theta
_{l}\right)  \text{,}\label{Et}\\
s\left(  t\right)   &  =\exp\left[  -\left(  t-T/2\right)  ^{2}/2\sigma
^{2}\right]  \text{,} \label{Shape}%
\end{align}
where $\left\{  \omega_{l}\right\}  $ are the allowed resonant transition
frequencies of the system, \textbf{and off-resonant excitation are not
included.} The controls are the amplitudes $\left\{  A_{l}\right\}  $ and
phases $\left\{  \theta_{l}\right\}  $, and the control interacts with the
system through the dipole operator $\mu$. In the laboratory such\textbf{
}control fields may be generated with programmable adaptive phase and
amplitude femtosecond pulse shaping techniques\cite{Nelson932032,
Weiner001929}\textbf{.}

In Eq.(\ref{rhos}) $\mathcal{F}\left\{  \rho_{S}(t)\right\}  $ is a functional
which represents decoherence caused by the environment and $\gamma$ is a
positive coefficient which indicates the strength of the decoherence and will
be varied in this paper to study the effect of dissipation on the control of
quantum dynamics. Various equations have been
developed\cite{Lindblad76119,Yan982721,Shuang002068} to describe the
interaction between a system and the environment. In this paper we take the
Lindblad form\cite{Lindblad76119}:%

\end{subequations}
\begin{equation}
\mathcal{F}\left\{  \rho_{S}(t)\right\}  =\sum_{\left\{  jn\right\}  }\left(
\mathcal{L}_{\left\{  jn\right\}  }\rho_{S}(t)\mathcal{L}_{\left\{
jn\right\}  }^{\dagger}-\frac{1}{2}\rho_{S}(t)\mathcal{L}_{\left\{
jn\right\}  }^{\dagger}\mathcal{L}_{\left\{  jn\right\}  }-\frac{1}%
{2}\mathcal{L}_{\left\{  jn\right\}  }^{\dagger}\mathcal{L}_{\left\{
jn\right\}  }\rho_{S}(t)\right)  \text{,} \label{LindbladF}%
\end{equation}
Here $\mathcal{L}_{\left\{  jn\right\}  }$ are bounded linear Lindblad
operators acting on $\rho_{S}$ such that the norm of the operators,
$\sum_{\left\{  jn\right\}  }\mathcal{L}_{\left\{  jn\right\}  }^{\dagger
}\mathcal{L}_{\left\{  jn\right\}  },$ is finite. In the Lindblad equation,
the Markovian and \textbf{weak coupling approximations are} made, and the
normalization and positive definite nature of the reduced density matrix
$\rho_{S}(t)$ is preserved.

In the model system studied here, the Lindblad operators in
Eq.(\ref{LindbladF}) are expressed phenomenologically as\cite{Gardiner2000}%

\begin{equation}
\mathcal{L}_{jn}=\sqrt{\Gamma_{jn}}\left\vert j\right\rangle \left\langle
n\right\vert ,\ j,n=0,1,...,N-1\text{,} \label{Ljn}%
\end{equation}
where $N$ is the number of levels and $\left\vert j\right\rangle $ are
eigenstates of the isolated system Hamiltonian $H_{0}$. The coefficient
$\Gamma_{jn}$, when multiplied by $\gamma$, specifies the transition rate
between level $j$ and $n$. These rates depend on how the system and
environment interact and will be chosen arbitrarily for the purposes of the
general analysis here.

Inserting Eq.(\ref{Ljn}) into Eq.(\ref{LindbladF}) produces the dynamical
equation for the reduced density matrix of the multi-level system,
\begin{subequations}
\label{Liou2}%
\begin{align}
\frac{\partial\rho_{ll^{\prime}}(t)}{\partial t}  &  =-i(\varepsilon
_{l}-\varepsilon_{l^{\prime}})\rho_{ll^{\prime}}(t)-i\sum_{n=0}^{N-1}\left(
\mu_{ln}\rho_{nl^{\prime}}(t)-\rho_{ln}(t)\mu_{nl^{\prime}}\right)
E(t)+\gamma\mathcal{F}\left\{  \rho_{S}(t)\right\}  _{ll^{\prime}}%
\text{,}\label{Lindblad}\\
\mathcal{F}\left\{  \rho_{S}(t)\right\}  _{ll^{\prime}}  &  =\delta
_{ll^{\prime}}\sum_{n=0}^{N-1}\Gamma_{ln}\rho_{nn}(t)-\frac{1}{2}\sum
_{n=0}^{N-1}\left(  \Gamma_{nl}+\Gamma_{nl^{\prime}}\right)  \rho_{ll^{\prime
}}(t)\text{.} \label{Fll}%
\end{align}
Closed-loop control simulations will be performed with model systems, but
under the standard cost function\cite{JM0010841} realizable in laboratory circumstances:%

\end{subequations}
\begin{subequations}
\label{Obj}%
\begin{align}
J\left[  E(t)\right]   &  =\left\vert O\left[  E(t),\gamma\right]
-O_{T}\right\vert ^{2}+\alpha F\text{,}\label{J0}\\
F  &  =\sum_{l}\left(  A_{l}\right)  ^{2}\text{,} \label{F0}%
\end{align}
where $O_{T}$ is the target value (expressed as a percent yield) and
\end{subequations}
\begin{equation}
O\left[  E(t),\gamma\right]  =\text{Tr}[\rho(T_{f})O_{f}] \label{O}%
\end{equation}
is the outcome produced by the trial field $E(t)$ under decoherence strength
$\gamma$, and $F$ is the fluence of the control field whose contribution is
weighted by the constant, $\alpha>0$. In the present work, $O_{f}=\left\vert
\Psi_{f}\right\rangle \left\langle \Psi_{f}\right\vert $ is a projection
operator for the population in a target state $\left\vert \Psi_{f}%
\right\rangle $. \bigskip

\section{Numerical Simulations}

To demonstrate how quantum optimal control can either fight or cooperate with
decoherence, we perform four simulations with simple model systems. The first
three simulations use the single path system in Figure (1a) which represents a
model multi-level system\cite{Shore90} or a truncated nonlinear
oscillator\cite{Larsen76254, Wallraff04431,Dykman2005_1}\textbf{,} while the
last simulation uses the double-path system in Figure (1b). With the single
path system, the first two models employ different decoherence coefficients
$\Gamma_{ij}$ associated with nearest neighbor transitions in Eq.(\ref{Fll}).
The third model allows as well for next nearest neighbor transitions. The
fourth model with the double-path system shows how cooperation between
decoherence and the control field can influence the transition path taken by
the dynamics. The decoherence term doesn't include spontaneous emission. All
of the simulations employ closed-loop optimization with a Genetic
Algorithm\cite{Goldberg97} (GA), in keeping with common laboratory practice.
Equation (\ref{J0}) is the cost function used to guide the GA determination of
the control field. The model parameters below are chosen to be illustrative of
the controlled physical phenomena, and similar behavior was found for other
choices as well.

\subsection{\bigskip Model 1}

This model uses the simple 5-level system in Figure (1a) with eigenstates
$\left\vert i\right\rangle $, $i=0,\cdots,4$ of the field free Hamiltonian
$H_{0}$, having only nearest neighbor transitions with the frequencies
$\omega_{01}=1.511$, $\omega_{12}=1.181$, $\omega_{23}=0.761$, $\omega
_{34}=0.553$ in rad$\cdot$fs$^{-1}$, and associated transition dipole moments
$\mu_{01}=0.5855$, $\mu_{12}=0.7079$, $\mu_{23}=0.8352$, $\mu_{34}=0.9281$ in
$10^{-30}$ C$\cdot$m. The decoherence coefficients $\Gamma_{ij}$ in
Eq.(\ref{Liou2}) are also non-zero only between nearest neighbor levels:
$\Gamma_{01}=\Gamma_{10}=0.0895$, $\Gamma_{12}=\Gamma_{21}=0.1942$,
$\Gamma_{23}=\Gamma_{32}=0.1209$, $\Gamma_{34}=\Gamma_{43}=0.2344$. The target
time is $T=200$ fs, and the pulse width in Eq.(\ref{Shape}) is $\sigma=30$ fs.
The control objective is to transfer population from the initially prepared
ground state $\left\vert 0\right\rangle $ to the final state $\left\vert
4\right\rangle $, such that $O=\left\vert 4\right\rangle \left\langle
4\right\vert $. In order to calibrate the strength of the decoherence, Table I
shows the state population distributions when the system is only driven by
various environmental interaction decoherence strengths. When $\gamma
\geqslant0.03$ fs$^{-1}$, decoherence is very strong as it drives more than
30\% of the population out of ground state at $t=200$ fs.

To examine the influence of decoherence when seeking an optimal control field,
first consider a high target yield (i.e., perfect control with the target
yield set at $O_{T}=100\%$ in Eq.(\ref{J0})). The simulation results in Table
II show that decoherence is always deleterious for achieving this target, but
a good yield is still possible. In order to reveal the separate contributions
of decoherence and the control field, Table II also shows the yield from the
decoherence alone $O\left[  E(t)=0,\gamma\right]  $ without the control field
being present and the yield from the control field alone $O\left[
E(t),\gamma=0\right]  $ without decoherence being present.

If we accept a low control yield outcome, very different control behavior is
found in the presence of decoherence. The results from optimizing
Eq.(\ref{J0}) are shown in Table III with various levels of decoherence for a
target yield of $O_{T}=5.0\%$. The results in Table III clearly indicate that
there is cooperation between the control field and the decoherence. For
example, when $\gamma=0.03$ fs$^{-1}$, the outcome of the control field and
decoherence together ($O[E(t),\gamma]=4.94\%$) is much larger than outcome
from control field alone ($O\left[  E(t),\gamma=0\right]  =0.63\%$) or the
decoherence alone ($O\left[  E(t)=0,\gamma\right]  =0.67\%$), and even larger
than their sum ($1.30\%$). Table III also shows that the control process
become more efficient with increasing decoherence strength ($\gamma$) as the
fluence of the control field is reduced while driving the dynamics to the same
target. The control field spectra at different decoherence strengths are
depicted in Fig.2. The mechanism of cooperation between the control field and
decoherence is not easy to identify from Fig.2. In the following simple model
the decoherence coefficients $\Gamma_{ij}$ are specially selected to identify
the mechanism of cooperation.

\subsection{\bigskip Model 2}

This model is similar with model 1. The only difference is that the transition
rates $\Gamma_{ij}$ in Eq.(\ref{Lindblad}) are chosen to increase
monotonically: $\Gamma_{01}=\Gamma_{10}=0.03495$, $\Gamma_{12}=\Gamma
_{21}=0.1242$, $\Gamma_{23}=\Gamma_{32}=0.3909$, $\Gamma_{34}=\Gamma
_{43}=0.6344$. Table IV is the low target yield analog of Table III,\ but for
model 2. Once again clear evidence is found for the optimal field cooperating
with the decoherence. In order to deduce the mechanism of cooperation between
the control field and the decoherence, Fig. 3 shows optimal control field
spectra at different levels of decoherence. The control field consists of four
peaks, corresponding to the four transitions between nearest neighbor levels.
With increasing decoherence strength, $\gamma$, the peak intensities
corresponding to transitions among the higher levels, become smaller, and
successively disappear. \textbf{This behavior shows that control field can be
found to cooperate with decoherence and drive the dynamics more efficiently.}
As a result, the cooperation between the control field and environment becomes
dramatic for $\gamma>0.03$ fs$^{-1}$. In this region the optimal field has
essentially no amplitude at the $2\rightarrow3$ and $3\rightarrow4$ transition
frequencies such that the field alone produces a vanishingly small yield, yet
with the environment present the cooperation is very efficient. If we chose
the decoherence coefficients $\Gamma_{ij}$ in Eq.(\ref{Lindblad}) decrease
monotonically(not shown here), then with increasing decoherence strength, the
peak intensities corresponding to transitions between the lower levels
disappear successively. This behavior indicates a similar cooperation
mechanism, but with a shift in the role of the various transitions.

\subsection{Model 3}

Model 3 is similar to model 1 except that two-quanta transitions are also
allowed: $\omega_{02}=2.692$, $\omega_{13}=1.942$, $\omega_{24}=1.314$ in
rad$\cdot$fs$^{-1}$, with transition dipole elements: $\mu_{02}=-0.1079$,
$\mu_{13}=-0.1823$, $\mu_{24}=-0.2786$ in 10$^{-3}$ C$\cdot$m. We also assume
that there are environmentally induced two-quanta transitions: $\Gamma
_{02}=0.01099$, $\Gamma_{13}=0.1087$, $\Gamma_{24}=0.1346$. Table V shows the
outcome from model 3 with a low yield target $O_{T}=10\%$. Once again,
cooperation between decoherence and the control field is found, although to a
lesser degree than in the earlier cases. The decrease in the degree of
cooperation appears mainly to arise from the enhanced target value of
$O_{T}=10\%$. As shown in model 1, eventually decoherence has a deleterious
effect when aiming towards a sufficiently high yield. However, under special
circumstances, such as decoherence breaking the symmetry of the Hamiltonian
$H(t)$, decoherence could still have a beneficial role when seeking a high yield.

\subsection{Model 4}

Model \ 4 is the more complex two-path system in Figure 1(b). In this model,
population can be transferred to the target state $\left\vert 4\right\rangle $
along two separate pathways. The transition frequencies, decoherence
coefficients and dipoles of the left path are the same as that of model 1; the
right path has the distinct transition frequencies: $\omega_{01^{\prime}%
}=2.513,\omega_{1^{\prime}2^{\prime}}=1.346,\omega_{2^{\prime}3^{\prime}%
}=0.345,\omega_{3^{\prime}4}=0.162$ in rad$\cdot$fs$^{-1}$, decoherence
coefficients: $\Gamma_{01^{\prime}}=\Gamma_{1^{\prime}0}=0.1164$,
$\Gamma_{1^{\prime}2^{\prime}}=\Gamma_{2^{\prime}1^{\prime}}=0.0885$,
$\Gamma_{2^{\prime}3^{\prime}}=\Gamma_{3^{\prime}2^{\prime}}=0.1557$,
$\Gamma_{3^{\prime}4}=\Gamma_{43^{\prime}}=0.1280$ and dipole elements:
$\mu_{01^{\prime}}=0.6525,\mu_{1^{\prime}2^{\prime}}=0.7848,\mu_{2^{\prime
}3^{\prime}}=0.9023,\mu_{3^{\prime}4}=1.0322$ in $10^{-30}$ C$\cdot$m. For
simplicity, only single quanta transitions are allowed along both paths.

The results in Table VI show cooperation between the field and the decoherence
in model 4 for a low target yield of $O_{T}=5\%$. Figure 4 shows the power
spectra of the three optimal control fields in Table VI. Panel (a) indicates
that in the case of no decoherence the control field primarily drives the
system along the right path. When decoherence is present in the left path
(panel (b)), the optimal control field chooses to drive the system dynamics
along the left path in order to cooperate with the decoherence. The lack of a
peak at $\omega_{34}$ in the latter case reflects the cooperation between the
control field and the environment. Similarly, in panel (c), when decoherence
is in the right path, the optimal field chooses to drive the system along the
right path. These results clearly show that efficiently achieving the yield is
the guiding principle dictating the nature of the control field and the
mechanism in these cases. The cost function in Eq.(\ref{Obj}) explicitly
contains a fluence term which naturally guides the closed-loop control search
towards efficient fields, and cooperating with decoherence is consistent with
this goal. The common circumstance in the laboratory of having a fixed maximum
fluence would produce similar behavior toward cooperating with decoherence
when it's beneficial.

\section{Foundations for Cooperating with Decoherence}

To illustrate the principle of how the control field can cooperate with
decoherence consider excitation along a ladder (or chain) of nondegenerate
transitions and energy levels, each linked only to its nearest neighbors as in
models 1-2 in section III. One can also think of this system as a truncated
nonlinear oscillator\cite{Larsen76254, Wallraff04431,Dykman2005_1}. The $N+1$
level system consists of an initially occupied ground state $\left\vert
0\right\rangle $, $N-1$ intermediate states $\left\vert n\right\rangle $, and
a final target state $\left\vert N\right\rangle $. The states are coupled with
an external laser pulse having the nominal form in Eq.(\ref{Et}). Both the
control field and the decoherence are assumed to be weak in this section,
corresponding to the low target yield cases in Section III.

Just as the parameter $\gamma$ characterizes the strength of decoherence, it
is convenient to introduce a parameter $\lambda$ to characterize the strength
of control field in Eq.(\ref{rhos}):
\begin{equation}
\frac{\partial\rho_{S}(t)}{\partial t}=-i[H_{0}-\mu\lambda E(t),\rho
_{S}(t)]+\gamma\mathcal{F}\left\{  \rho_{S}(t)\right\}  \text{.}%
\end{equation}
In the analysis below we shall focus on the regime $\lambda\rightarrow0$ and
$\gamma\rightarrow0$. Transforming the density matrix $\rho_{S}\left(
t\right)  $ to the interaction picture $\rho\left(  t\right)  =e^{-iH_{0}%
t}\rho_{S}\left(  t\right)  e^{iH_{0}t}$, or equivalently
\begin{equation}
\rho_{ll^{\prime}}=\left(  \rho_{S}\right)  _{ll^{\prime}}e^{-i\left(
\varepsilon_{l}-\varepsilon_{l^{\prime}}\right)  t} \label{RhoI}%
\end{equation}
produces%
\begin{subequations}
\begin{align}
\frac{\partial}{\partial t}\rho\left(  t\right)   &  =\left[  i\lambda
\mathcal{L}_{I}\left(  t\right)  +\gamma\mathcal{F}\right]  \rho\left(
t\right)  \text{,}\label{Lint}\\
\mathcal{L}_{I}\left(  t\right)  \rho\left(  t\right)   &  =\left[
\mu^{\left(  I\right)  }(t)E(t),\rho\left(  t\right)  \right]  \text{,}%
\end{align}
where $\mu^{\left(  I\right)  }\left(  t\right)  =e^{-iH_{0}t}\mu e^{iH_{0}t}%
$, or in matrix form $\mu_{kj}^{\left(  I\right)  }\left(  t\right)  =\mu
_{kj}e^{-i\omega_{kj}t}$ with $\omega_{kj}=\varepsilon_{j}-\varepsilon_{k}$.
It is easy to show that the decoherence functional $\mathcal{F}$ defined in
Eq.(\ref{Fll}) is not changed under the transformation of Eq.(\ref{RhoI}). The
solution of Eq.(\ref{Lint}) can be expressed formally as
\end{subequations}
\begin{equation}
\rho\left(  T_{f}\right)  =\exp_{+}\left[  \int_{0}^{T_{f}}\left(
i\lambda\mathcal{L}_{I}\left(  t\right)  +\gamma\mathcal{F}\right)  dt\right]
\rho\left(  0\right)  \text{,}%
\end{equation}
where $\exp_{+}$ is the time-ordering exponential operator. In order to
explore the scaling with $\gamma$ and $\lambda$ we shall now introduce the
first order Magnus approximation\cite{Magnus54649}, which ignores the time
ordering , such that%
\begin{subequations}
\begin{align}
\rho\left(  T_{f}\right)   &  \approx\exp\left[  \int_{0}^{T_{f}}\left(
i\lambda\mathcal{L}_{I}\left(  t\right)  +\gamma\mathcal{F}\right)  dt\right]
\rho\left(  0\right) \label{M1}\\
&  =\exp\left[  \left(  i\lambda\mathcal{E}+\gamma\mathcal{F}\right)
T_{f}\right]  \rho\left(  0\right)  \text{,} \label{M2}%
\end{align}
where $\mathcal{E}$ is a time-independent functional acting as%
\end{subequations}
\begin{equation}
\mathcal{E}\rho=\left[  W,\rho\right]
\end{equation}
with the matrix $W$ defined as
\begin{subequations}
\begin{align}
W  &  =\frac{1}{T_{f}}\int_{0}^{T_{f}}\mu^{\left(  I\right)  }\left(
t\right)  E\left(  t\right)  dt\text{,}\label{U}\\
W_{kj}  &  =\frac{1}{T_{f}}\int_{0}^{T_{f}}\mu_{kj}e^{-i\omega_{kj}t}E\left(
t\right)  dt\approx\mu_{kj}\frac{\varepsilon\left(  \omega_{kj}\right)
}{T_{f}}\text{.} \label{Uij}%
\end{align}
Here $\epsilon\left(  \omega\right)  $ denotes the spectrum of the control
pulse:%
\end{subequations}
\begin{equation}
\epsilon\left(  \omega\right)  =\int_{-\infty}^{\infty}E(t)e^{-i\omega
t}dt=\sum_{l=1}^{M}A_{l}\left[  g\left(  \omega-\omega_{l}\right)
e^{i\theta_{l}}+g\left(  \omega+\omega_{l}\right)  e^{-i\theta_{l}}\right]
\text{,} \label{ew}%
\end{equation}
with $g\left(  \omega\right)  $ being the Fourier transform of the shape
function $s(t)$ in Eq.(\ref{Shape}). The first order Magnus approximation
suffices in the "impulse limit" of a sufficiently short
pulse\cite{Light663897}.

It's convenient to introduce double bracket notation\cite{Mukamel1995} where
the ket $\left\vert \left.  jk\right\rangle \right\rangle $ denotes the
Liouville space vector representing the Hilbert space operator $\left\vert
j\right\rangle \left\langle k\right\vert $ and the bra $\left\langle
\left\langle jk\right.  \right\vert $ as the Hermitian conjugate to
$\left\vert \left.  jk\right\rangle \right\rangle $: $\left\langle
\left\langle jk\right.  \right\vert \equiv\left(  \left\vert \left.
jk\right\rangle \right\rangle \right)  ^{\dagger}\equiv\left\vert
k\right\rangle \left\langle j\right\vert $. From Eqs.(\ref{O}) and (\ref{M2}),
the outcome from applying the control field can be expanded and rewritten as%
\begin{subequations}
\begin{align}
O\left[  E\left(  t\right)  ,\gamma\right]   &  =\left\langle \left\langle
NN\right.  \right\vert \exp\left[  \left(  i\lambda\mathcal{E}+\gamma
\mathcal{F}\right)  T_{f}\right]  \left\vert \left.  00\right\rangle
\right\rangle \label{OEg}\\
&  =\sum_{k=0}^{\infty}\left\langle \left\langle NN\right.  \right\vert
\left(  i\lambda\mathcal{E}+\gamma\mathcal{F}\right)  ^{k}\mathbf{P}%
^{k}\left(  T_{f}\right)  \left\vert \left.  00\right\rangle \right\rangle
\text{,} \label{OS}%
\end{align}
where the functional $\mathbf{P}$ is defined by:%
\end{subequations}
\begin{subequations}
\begin{align}
\mathbf{P}\left[  f\right]  \left(  t\right)   &  =\int_{0}^{t}f\left(
t_{1}\right)  dt_{1}\\
\mathbf{P}^{k}\left[  f\right]  \left(  t\right)   &  =\int_{0}^{t}dt_{1}%
\int_{0}^{t_{1}}dt_{2}\cdots\int_{0}^{t_{k-1}}dt_{k}f\left(  t_{k}\right)
\text{.}%
\end{align}
It is easy to verify that $P^{k}\left[  f\right]  \left(  t\right)
=\frac{t^{k}}{k!}$ for the identity function $f(t)\equiv1$.

\subsection{Dynamics Driven by the Control Field Alone}

When there is no decoherence, the yield from the control field is
\end{subequations}
\begin{equation}
O\left[  E\left(  t\right)  ,0\right]  =\sum_{k=0}^{\infty}\frac{T_{f}^{k}%
}{k!}\left\langle \left\langle NN\right.  \right\vert \left(  i\mathcal{E}%
\right)  ^{k}\left\vert \left.  00\right\rangle \right\rangle \lambda
^{k}\text{.} \label{OEL}%
\end{equation}
The system yield can also be described in Hilbert space by%
\begin{equation}
O\left[  E\left(  t\right)  ,0\right]  =\left\vert \left\langle N\right\vert
e^{iWT_{f}\lambda}\left\vert 0\right\rangle \right\vert ^{2}=\left\vert
\sum_{k=0}^{\infty}\frac{T_{f}^{k}}{k!}\left\langle N\right\vert \left(
iW\right)  ^{N}\left\vert 0\right\rangle \lambda^{k}\right\vert ^{2}\text{.}
\label{OEH}%
\end{equation}
For a ladder system, $W$ is a tridiagonal matrix. When $\lambda\rightarrow0$,
we only need to keep the first non-zero terms of Eq.(\ref{OEH}),%
\begin{equation}
O\left[  E\left(  t\right)  ,0\right]  \approx\left\vert \frac{T_{f}^{N}}%
{N!}\left\langle N\right\vert \left(  iW\right)  ^{N}\left\vert 0\right\rangle
\right\vert ^{2}\lambda^{2N}=\frac{T_{f}^{2N}}{N!^{2}}T_{0N}^{2}\lambda
^{2N}\text{,}%
\end{equation}
where the coherent transition elements $T_{mn}$ are defined as
\begin{subequations}
\begin{align}
T_{mn}  &  =\left\vert \left\langle m\right\vert W^{n-m}\left\vert
n\right\rangle \right\vert =%
%TCIMACRO{\dprod \limits_{k=m+1}^{n}}%
%BeginExpansion
{\displaystyle\prod\limits_{k=m+1}^{n}}
%EndExpansion
\left\vert W_{k,k-1}\right\vert =%
%TCIMACRO{\dprod \limits_{k=n}^{m}}%
%BeginExpansion
{\displaystyle\prod\limits_{k=n}^{m}}
%EndExpansion
\Omega_{k}\text{,}\\
\Omega_{k}  &  =\left\vert W_{k,k-1}\right\vert =\mu_{k,k-1}\frac{\left\vert
\varepsilon\left(  \omega_{k,k-1}\right)  \right\vert }{T_{f}}\text{.}%
\end{align}
By comparing Eqs.(\ref{OEL}) and (\ref{OEH}), we know that the first non-zero
term in Eq.(\ref{OEL}) is $k=2N$%
\end{subequations}
\begin{equation}
O\left[  E\left(  t\right)  ,0\right]  \approx\frac{T_{f}^{2N}}{2N!}%
\left\langle \left\langle NN\right.  \right\vert \left(  i\mathcal{E}\right)
^{2N}\left\vert \left.  00\right\rangle \right\rangle \lambda^{2N}\text{,}%
\end{equation}
and the following equality holds:%
\begin{equation}
\left\langle \left\langle NN\right.  \right\vert \mathcal{E}^{2N}\left\vert
\left.  00\right\rangle \right\rangle =C_{2N}^{N}T_{0N}^{2}\text{.}%
\end{equation}
The extension of the above equality to additional matrix elements is
straightforward,%
\begin{equation}
\left\langle \left\langle n+m,n+m\right.  \right\vert \left(  i\mathcal{E}%
\right)  ^{2m}\left\vert \left.  nn\right\rangle \right\rangle =C_{2m}%
^{m}T_{n,n+m}^{2}\text{.} \label{LtoH}%
\end{equation}
Finally, the lowest order non-zero outcome from applying the control field
without decoherence is
\begin{equation}
O\left[  E\left(  t\right)  ,0\right]  =\lambda^{2N}\frac{T_{f}^{2N}}{N!^{2}%
}T_{0N}^{2}\text{.} \label{OEFinal}%
\end{equation}

\subsection{Dynamics Driven by Environmental Decoherence Alone}

For simplicity, the decoherence is assumed to only induce transitions between
nearest neighbor levels in the system, so that the decoherence matrix $\Gamma$
is tridiagonal with elements%
\begin{equation}
\Gamma_{ij}=\left\langle \left\langle ii\right.  \right\vert \mathcal{F}%
\left\vert \left.  jj\right\rangle \right\rangle \text{,}%
\end{equation}
and $\Gamma$ plays a similar role as the dipole matrix. Decoherence can also
drive the system from the initial state to the target state, at least to some
degree, when there is no external field. The outcome due to environmental
decoherence without the control field is%
\begin{equation}
O\left[  0,\gamma\right]  =\sum_{k=1}^{\infty}\frac{T_{f}^{k}}{k!}\left\langle
\left\langle NN\right.  \right\vert \mathcal{F}^{k}\left\vert \left.
00\right\rangle \right\rangle \gamma^{k}\text{.} \label{OD}%
\end{equation}
It is easy to obtain the following decoherence transition elements,%
\begin{equation}
\left\langle \left\langle n+m,n+m\right.  \right\vert \mathcal{F}%
^{m}\left\vert \left.  nn\right\rangle \right\rangle =\prod_{k=1}^{m}%
\gamma_{n+k}\text{,} \label{Fele}%
\end{equation}
where $\gamma_{k}$ is the $k$-th decoherence-induced transition rate between
the nearest neighbor levels,
\begin{equation}
\gamma_{k}=\Gamma_{k+1,k}\text{.}%
\end{equation}
If the decoherence coupling is weak, $\gamma\rightarrow0$, only the lowest
non-zero term of Eq.(\ref{OD}) is needed,%
\begin{subequations}
\begin{align}
O\left[  E\left(  t\right)  =0,\gamma\right]   &  \approx\frac{T_{f}^{N}}%
{N!}\left\langle \left\langle NN\right.  \right\vert \mathcal{F}^{N}\left\vert
\left.  00\right\rangle \right\rangle \gamma^{N}\label{Og}\\
&  =\gamma^{N}\frac{T_{f}^{N}}{N!}%
%TCIMACRO{\dprod \limits_{k=1}^{N}}%
%BeginExpansion
{\displaystyle\prod\limits_{k=1}^{N}}
%EndExpansion
\gamma_{k}\text{,} \label{ODFinal}%
\end{align}
which corresponds to a transition path $\left\vert \left.  00\right\rangle
\right\rangle \rightarrow\left\vert \left.  11\right\rangle \right\rangle
\rightarrow\cdots\rightarrow\left\vert \left.  NN\right\rangle \right\rangle $.

\subsection{Cooperation between the Control Field and Decoherence}

Consider the case where the control field and environmental decoherence are
present simultaneously, so that the outcome is described by Eq.(\ref{M2}). If
the control field and decoherence are both weak, the perturbation
approximation can be used again. Strong cooperation between the control field
(Eq.(\ref{OEFinal})) and decoherence (Eq.(\ref{ODFinal})) is expected when
their contributions to the outcome are the same order, which corresponds to
\end{subequations}
\begin{equation}
\gamma=\lambda^{2}\text{.}%
\end{equation}
Then the perturbative solution $\rho\left(  t\right)  $ has the property
\begin{equation}
\rho_{mn}\left(  t\right)  \sim\lambda^{m+n}\text{.}%
\end{equation}
$\rho\left(  T_{f}\right)  $ in Eq.(\ref{M2}) is the solution of following
equation
\begin{equation}
\frac{\partial\rho\left(  t\right)  }{\partial t}=\left(  i\lambda
\mathcal{E}+\lambda^{2}\mathcal{F}\right)  \rho\left(  t\right)  \text{,}%
\end{equation}
expressed explicitly as%
\begin{equation}
\frac{\partial\rho_{ll^{\prime}}}{\partial t}\approx i\lambda\left(
W_{l,l-1}\rho_{l-1,l^{\prime}}-\rho_{l,l^{\prime}-1}W_{l^{\prime}-1,l^{\prime
}}\right)  +\lambda^{2}\delta_{ll^{\prime}}\Gamma_{l,l-1}\rho_{l-1,l-1}%
\text{,} \label{PurDym}%
\end{equation}
where the terms higher than $\lambda^{l+l^{\prime}}$ are neglected. The last
term of Eq.(\ref{PurDym}) represents the effect of decoherence which only
drives transitions of the type $\left\vert \left.  l-1,l-1\right\rangle
\right\rangle \rightarrow\left\vert \left.  ll\right\rangle \right\rangle $.
If a high control yield is expected and the perturbation theory approximation
is invalid, all of the decoherence terms in Eq.(\ref{Fll}) have to be
considered, where some terms will interact destructively with the control
field induced dynamics,
\begin{equation}
\left(  \mathcal{F}\rho\right)  _{ll^{\prime}}=\delta_{ll^{\prime}}%
\Gamma_{l,l-1}\rho_{l-1,l-1}+\delta_{ll^{\prime}}\Gamma_{l,l+1}\rho
_{l+1,l+1}-\frac{1}{2}\left(  \Gamma_{l-1,l}+\Gamma_{l+1,l}+\Gamma_{l^{\prime
}-1,l^{\prime}}+\Gamma_{l^{\prime}+1,l^{\prime}}\right)  \rho_{ll^{\prime}%
}\text{.}%
\end{equation}
Here it's the first term that cooperates with the control field.

The outcome from both the control field and decoherence is
\begin{subequations}
\begin{align}
O\left[  E\left(  t\right)  ,\gamma\right]   &  =\left\langle \left\langle
NN\right.  \right\vert \exp\left[  \left(  \lambda i\mathcal{E}+\lambda
^{2}\mathcal{F}\right)  T_{f}\right]  \left\vert \left.  00\right\rangle
\right\rangle \\
&  =\sum_{k=1}^{\infty}\left\langle \left\langle NN\right.  \right\vert
\left[  \lambda i\mathcal{E}\mathbf{P}\left(  T_{f}\right)  +\lambda
^{2}\mathcal{F}\mathbf{P}\left(  T_{f}\right)  \right]  ^{k}\left\vert \left.
00\right\rangle \right\rangle \text{.}%
\end{align}
Expanding the above equation and keeping only terms of order $\lambda^{2N}$,
we get the perturbation approximation for the outcome,
\end{subequations}
\begin{equation}
O\left[  E\left(  t\right)  ,\gamma\right]  \approx\lambda^{2N}\sum_{m=0}%
^{N}\sum_{0\leq n_{1}<\cdots<n_{m}<N}A\left(  n_{1},n_{2},\cdots,n_{m}\right)
\text{,} \label{OEGFinal}%
\end{equation}
with $A\left(  n_{1},n_{2},\cdots,n_{m}\right)  $ given by%
\begin{multline}
A\left(  n_{1},n_{2},\cdots,n_{m}\right)  =\mathbf{P}^{2N-m}\left(
T_{f}\right)  \left\langle \left\langle NN\right.  \right\vert \left(
i\mathcal{E}\right)  ^{2\left(  N-n_{m}\right)  }\left\vert \left.  n_{m}%
n_{m}\right\rangle \right\rangle \label{A0}\\
\times%
%TCIMACRO{\dprod \limits_{k=1}^{m}}%
%BeginExpansion
{\displaystyle\prod\limits_{k=1}^{m}}
%EndExpansion
\left\langle \left\langle n_{k}n_{k}\right.  \right\vert \left(
i\mathcal{E}\right)  ^{2\left(  n_{k}-n_{k-1}-1\right)  }\left\vert \left.
n_{k-1}+1,n_{k-1}+1\right\rangle \right\rangle \left\langle \left\langle
n_{k-1}n_{k-1}\right.  \right\vert \mathcal{F}\left\vert \left.
n_{k-1}+1,n_{k-1}+1\right\rangle \right\rangle \text{,}%
\end{multline}
describing the contribution of the paths in which the decoherence operator
$\mathcal{F}$ drives the system from state $\left\vert \left.  n_{k}%
-1,n_{k}-1\right\rangle \right\rangle $ to $\left\vert \left.  n_{k}%
,n_{k}\right\rangle \right\rangle $, and the control field operator
$\mathcal{E}$ drives the system from state $\left\vert \left.  n_{k}%
+1,n_{k}+1\right\rangle \right\rangle $ to $\left\vert \left.  n_{k+1}%
n_{k+1}\right\rangle \right\rangle $. Substituting Eqs.(\ref{LtoH}) and
(\ref{Fele}) into Eq.(\ref{A0}), we have%
\begin{subequations}
\begin{align}
A\left(  n_{1},n_{2},\cdots,n_{m}\right)   &  =\frac{T_{f}^{2N-m}}{\left(
2N-m\right)  !}C_{2\left(  N-n_{m}-1\right)  }^{N-n_{m}-1}%
%TCIMACRO{\dprod \limits_{k=1}^{m}}%
%BeginExpansion
{\displaystyle\prod\limits_{k=1}^{m}}
%EndExpansion
T_{n_{k-1}+1,n_{k}}^{2}C_{2\left(  n_{k}-n_{k-1}-1\right)  }^{n_{k}-n_{k-1}%
-1}\label{A1}\\
&  =T_{0N}^{2}\frac{T_{f}^{2N-m}}{\left(  2N-m\right)  !}C_{2\left(
N-n_{m}-1\right)  }^{N-n_{m}-1}%
%TCIMACRO{\dprod \limits_{k=1}^{m}}%
%BeginExpansion
{\displaystyle\prod\limits_{k=1}^{m}}
%EndExpansion
\frac{\gamma_{n_{k}}}{\Omega_{n_{k}}^{2}}C_{2\left(  n_{k}-n_{k-1}-1\right)
}^{n_{k}-n_{k-1}-1}\text{.} \label{A2}%
\end{align}
\qquad If each term in Eq.(\ref{Et}) is only resonant with a single
transition,
\end{subequations}
\begin{equation}
\omega_{n}\approx\varepsilon_{n+1}-\varepsilon_{n}\text{,}%
\end{equation}
using the RWA\cite{Shore90} (Rotating Wave Approximation) we can drop all
non-resonant terms in Eq.(\ref{Lint}),
\begin{equation}
\mu^{\left(  I\right)  }(t)E(t)\approx s\left(  t\right)  H_{F}\text{.}%
\end{equation}
The time-independent $H_{F}$ matrix elements are%
\begin{equation}
\left(  H_{F}\right)  _{kj}=\mu_{j}A_{j}\delta_{k,j+1}+\mu_{k}A_{k}%
\delta_{k+1,j}\text{.}%
\end{equation}
Here, $\mu_{k}=\mu_{k,k-1}$ denotes the $k$-th dipole element between the
nearest levels. The operator $\mathcal{L}_{I}\left(  t\right)  $ commutes with
itself at different times implying that the Magnus approximation is valid in
the present circumstance under the RWA, and it's easy to get Eq.(\ref{A2})
with $\Omega_{k}$ being
\begin{equation}
\Omega_{k}=\mu_{k}A_{k}\frac{T_{e}}{T_{f}}\text{,}%
\end{equation}
defining the effective pulse duration as
\begin{equation}
T_{e}=\int_{0}^{T_{f}}s\left(  t\right)  dt\text{.}%
\end{equation}
It is also easy to show that the outcome of Eq.(\ref{OEGFinal}) can be
decomposed into a linear combination of contributions from the field component
intensity $A_{j}^{2}$ and the decoherence transition coefficients $\gamma_{j}$,%

\begin{equation}
O\left[  E\left(  t\right)  ,\gamma\right]  =A_{j}^{2}F_{1j}+\gamma_{j}%
F_{2j}\text{,} \label{Decomp}%
\end{equation}
where $F_{1j}$ and $F_{2j}$ are functions independent of $A_{j}$ and
$\gamma_{j}$, respectively. There is clearly cooperation between $A_{j}$\ (a
coherently driven transition) and $\gamma_{j}$ (a decoherently driven
transition). For example, the outcome for a two level system is%
\begin{equation}
\rho_{11}\left(  T_{f}\right)  =\left(  \mu_{1}A_{1}\right)  ^{2}T_{e}%
^{2}+\gamma_{1}T_{f}\text{,}%
\end{equation}
and the outcome for a three level system is%
\begin{equation}
\rho_{22}\left(  T_{f}\right)  =\frac{1}{4}\left(  \mu_{1}A_{1}\mu_{2}%
A_{2}\right)  ^{2}T_{e}^{4}+\frac{1}{3}\left(  \mu_{1}A_{1}\right)  ^{2}%
\gamma_{2}T_{e}^{2}T_{f}+\frac{1}{3}\left(  \mu_{2}A_{2}\right)  ^{2}%
\gamma_{1}T_{e}^{2}T_{f}+\frac{1}{2}\gamma_{1}\gamma_{2}T_{f}^{2}\text{.}%
\end{equation}
The objective cost function in Eq.(\ref{J0}) can be written in terms of the
contributions from each specific control field intensity $A_{j}^{2}$,%

\begin{equation}
J\left(  A_{j}\right)  =\left(  A_{j}^{2}F_{1j}+\gamma_{j}F_{2j}-O_{T}\right)
^{2}+\alpha A_{j}^{2}+\alpha\sum_{k\neq j}A_{k}^{2}\text{.} \label{JA}%
\end{equation}
Here, $F_{1j}$ and $F_{2j}$ are independent of both $A_{j}$ and $\gamma_{j}$.
It is easy to show that the condition for optimizing Eq.(\ref{JA}) with
respect to $A_{j}$ is%
\begin{equation}
A_{j}^{2}F_{1j}+\gamma_{j}F_{2j}=O_{T}-\frac{\alpha}{2F_{1j}}\text{,}%
\end{equation}
The latter equation indicates that the contribution from decoherence can
beneficially act to decrease the required amplitude of the optimal control
field to attain the same yield.

\section{Conclusions}

Various impacts of decoherence upon quantum control are explored in this
paper. Numerical simulations from several cases indicate that control fields
can be found that either cooperate with or fight against decoherence,
depending on the circumstances. Two extreme cases of high and low target yield
are chosen to illustrate distinct control behavior in the presence of
decoherence: (a) the control field can fight against decoherence effectively
when a high yield is desired, while (b) in the case of a low target yield, the
control field can even cooperate with decoherence to drive the dynamics while
minimizing the control fluence.

Four models were studied and a \textbf{first-order perturbation theory }of the
weak control and decoherence limitis presented in this paper. Models 1 and 2
considered the control of the same system in different decoherence
environments. Both cases show clear cooperation between the control field and
the decoherence driven dynamics for low target yield. The detailed mechanistic
role of decoherence can be subtle as indicated in model 1, although the final
physical impact may be simple to understand. Model 2 clearly identified the
role of decoherence by setting up a special structured interaction between the
system and the environment. Although the conclusions in this paper are based
on simple physical models with the Lindblad equation, similar behavior is
expected for other systems and formulations of decoherence. The findings in
this work are parallel with an analogous study\cite{Shuang049270} on the role
of control noise. These work aim to provide a better foundation to understand
the physical processes at work when using closed-loop optimization in the
laboratory, even in cases of significant noise and decoherence.

\begin{acknowledgments}
The author acknowledge support from the National Science Foundation and an ARO
MURI grant.
\end{acknowledgments}

\pagebreak

\bigskip Table I. Population distribution of the single path model 1 when the
system is only driven by interaction with the environment.

\noindent%
%TCIMACRO{\TeXButton{btab}{\begin{ruledtabular}}}%
%BeginExpansion
\begin{ruledtabular}%
%EndExpansion%
\begin{tabular}
[c]{cccccc}
& \multicolumn{5}{c}{Population in the state (\%)}\\\cline{2-6}%
$\gamma\ ^{a}$(fs$^{-1}$) & 0 & 1 & 2 & 3 & 4\\\hline
0.05 & 52.8 & 25.5 & 14.6 & 4.64 & 2.39\\
0.03 & 65.0 & 22.7 & 9.65 & 1.98 & 0.67\\
0.01 & 84.8 & 12.7 & 2.25 & 0.17 & 0.02\\
0.00 & 100 & 0 & 0 & 0 & 0
\end{tabular}%
%TCIMACRO{\TeXButton{etab}{\end{ruledtabular}}}%
%BeginExpansion
\end{ruledtabular}%
%EndExpansion

$^{a}$ Strength of decoherence, refer to Eqs.(\ref{Liou1}) and Eq.(\ref{Liou2})

\bigskip

\bigskip

Table II. Optimal control fields fighting against decoherence with model 1 for
the highest possible objective yield of $O_{T}=100\%$.

\noindent%
%TCIMACRO{\TeXButton{btab}{\begin{ruledtabular}}}%
%BeginExpansion
\begin{ruledtabular}%
%EndExpansion%
\begin{tabular}
[c]{cccccc}%
$\gamma$(fs$^{-1}$) & $O\left[  E(t)=0,\gamma\right]  ^{a}$(\%) & $O\left[
E^{op}(t),\gamma=0\right]  ^{b}$(\%) & $O\left[  E^{op}(t),\gamma\right]
^{c}$(\%) & $O\left[  E_{0}^{op}(t),\gamma\right]  ^{e}$(\%) & $F\;^{d}%
$\\\hline
0.05 & 2.39 & 97.58 & 34.90 & 34.58 & 0.071\\
0.03 & 0.67 & 98.43 & 46.61 & 46.42 & 0.067\\
0.01 & 0.02 & 98.65 & 72.90 & 72.79 & 0.066\\
0.00 & 0.00 & 98.53 & 98.53 & 98.53 & 0.064
\end{tabular}%
%TCIMACRO{\TeXButton{etab}{\end{ruledtabular}}}%
%BeginExpansion
\end{ruledtabular}%
%EndExpansion

$^{a}$ Yield from decoherence alone without a control field

$^{b}$ Yield arising from the control field without decoherence, but the
control field is determined in the presence of decoherence at the specified
value of $\gamma$.

$^{c}$ Yield from the the optimal control field in the presence of
decoherence, Eq.(\ref{O}).

$^{d}$ Fluence of the control field.

$^{e}$ Yield from decoherence and the field optimized with zero decoherence.

\bigskip\pagebreak

\bigskip Table III. Yields from optimal control fields with different levels
of decoherence in the case of model 1 for a low objective yield of
$O_{T}=5.0\%$.

\noindent%
%TCIMACRO{\TeXButton{btab}{\begin{ruledtabular}}}%
%BeginExpansion
\begin{ruledtabular}%
%EndExpansion%
\begin{tabular}
[c]{ccccc}%
$\gamma$(fs$^{-1}$) & $O\left[  E(t)=0,\gamma\right]  ^{a}$(\%) & $O\left[
E^{op}(t),\gamma=0\right]  ^{b}$(\%) & $O\left[  E^{op}(t),\gamma\right]
^{c}$(\%) & $F\;^{d}$\\\hline
0.05 & 2.39 & 2$\times10^{-4}$ & 4.92 & 4.08$\times10^{-3}$\\
0.03 & 0.67 & 0.63 & 4.94 & 8.17$\times10^{-3}$\\
0.01 & 0.02 & 3.01 & 4.96 & 1.23$\times10^{-2}$\\
0.00 & 0.0 & 4.96 & 4.96 & 1.39$\times10^{-2}$%
\end{tabular}%
%TCIMACRO{\TeXButton{etab}{\end{ruledtabular}}}%
%BeginExpansion
\end{ruledtabular}%
%EndExpansion

\bigskip$^{a,b,c,d}$ refer to Table II.

\bigskip

\bigskip

Table IV. Optimal control with decoherence for model 2 with a low objective
yield of $O_{T}=5.0\%$.

\noindent%
%TCIMACRO{\TeXButton{btab}{\begin{ruledtabular}}}%
%BeginExpansion
\begin{ruledtabular}%
%EndExpansion%
\begin{tabular}
[c]{ccccc}%
$\gamma$(fs$^{-1}$) & $O\left[  E(t)=0,\gamma\right]  ^{a}$(\%) & $O\left[
E^{op}(t),\gamma=0\right]  ^{b}$(\%) & $O\left[  E^{op}(t),\gamma\right]
^{c}$(\%) & $F\;^{d}$\\\hline
0.05 & 2.2 & 7.69$\times10^{-12}$ & 4.98 & 2.07$\times10^{-3}$\\
0.03 & 0.71 & 2.51$\times10^{-9}$ & 5.08 & 5.94$\times10^{-3}$\\
0.01 & 0.03 & 0.52 & 4.90 & 1.23$\times10^{-2}$\\
0.00 & 0.0 & 4.96 & 4.96 & 1.40$\times10^{-2}$%
\end{tabular}%
%TCIMACRO{\TeXButton{etab}{\end{ruledtabular}}}%
%BeginExpansion
\end{ruledtabular}%
%EndExpansion

$^{a,b,c,d}$ refer to Table II.

\pagebreak

\bigskip Table V. Optimal control with decoherence for model 3 with a low
objective yield of $O_{T}=10.0\%$.

\noindent%
%TCIMACRO{\TeXButton{btab}{\begin{ruledtabular}}}%
%BeginExpansion
\begin{ruledtabular}%
%EndExpansion%
\begin{tabular}
[c]{ccccc}%
$\gamma$(fs$^{-1}$) & $O\left[  E(t)=0,\gamma\right]  ^{a}$(\%) & $O\left[
E^{op}(t),\gamma=0\right]  ^{b}$(\%) & $O\left[  E^{op}(t),\gamma\right]
^{c}$(\%) & $F\;^{d}$\\\hline
0.05 & 5.24 & 3.49 & 9.92 & 1.03$\times10^{-2}$\\
0.03 & 2.08 & 4.73 & 10.00 & 1.17$\times10^{-2}$\\
0.01 & 0.19 & 10.06 & 10.00 & 1.84$\times10^{-2}$\\
0.00 & 0.0 & 10.00 & 10.00 & 1.70$\times10^{-2}$%
\end{tabular}%
%TCIMACRO{\TeXButton{etab}{\end{ruledtabular}}}%
%BeginExpansion
\end{ruledtabular}%
%EndExpansion

\bigskip$^{a,b,c,d}$ refer to Table II.

\bigskip

\bigskip

Table VI. Yield attained from the optimal field for model 4 with the low
objective yield of $O_{T}=5\%$.

\noindent%
%TCIMACRO{\TeXButton{btab}{\begin{ruledtabular}}}%
%BeginExpansion
\begin{ruledtabular}%
%EndExpansion%
\begin{tabular}
[c]{ccccc}%
$\gamma=(\gamma_{L},\gamma_{R})^{e}$(fs$^{-1}$) & $\left\langle O\left[
E(t)=0,\gamma\right]  \right\rangle _{N}{}^{a}$(\%) & $O\left[  E^{op}%
(t),\gamma=0\right]  ^{b}$(\%) & $\left\langle O\left[  E^{op}(t),\gamma
\right]  \right\rangle ^{c}$(\%) & $F\ ^{d}$\\\hline
0.00,0,00 & 0.00 & 4.99 & 4.99 & 1.18$\times10^{-2}$\\
0.04,0.00 & 1.42 & 1.34$\times10^{-4}$ & 4.92 & 6.20$\times10^{-3}$\\
0.00,0.04 & 0.85 & 0.33 & 4.95 & 5.98$\times10^{-3}$%
\end{tabular}%
%TCIMACRO{\TeXButton{etab}{\end{ruledtabular}}}%
%BeginExpansion
\end{ruledtabular}%
%EndExpansion

\bigskip$^{a,b,c,d}$ refer to Table II.

$^{e}$\ $\gamma_{L}$,$\gamma_{R}$: Decoherence strength in the left and right
paths, respectively;

\pagebreak

\noindent%

%TCIMACRO{\FRAME{ftbpFU}{4.1174in}{3.0885in}{0pt}{\Qcb{Two multilevel quantum
%systems used to investigate the impact of decoherence on optimally controlled
%dynamics. (a) The five level single path system used for simulations with
%models 1--3 in Sections III.A--III.C, respectively. Models 1 and 2 only allow
%for single quanta transitions (solid arrows) while model 3 allows both single
%and double quanta transitions (dashed arrows). (b) The double-path system for
%model 4 in Section 3.D.\bigskip}}{}{fig1.eps}%
%width 4.1174in;  height 3.0885in;  depth 0pt;  original-width 3.5754in;
%original-height 2.9022in;  cropleft "0";  croptop "1";  cropright "1.0849";
%cropbottom "0";  filename 'Fig1.eps';file-properties "XNPEU";}}}%
%BeginExpansion
\begin{figure}
[ptb]
\begin{center}
\includegraphics[
trim=0.000000in 0.000000in -0.303552in 0.000000in,
height=3.0885in,
width=4.1174in
]%
{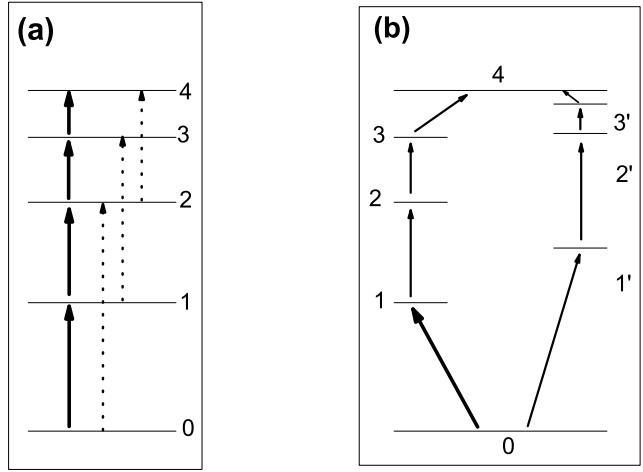}%
\caption{Two multilevel quantum systems used to investigate the impact of
decoherence on optimally controlled dynamics. (a) The five level single path
system used for simulations with models 1--3 in Sections III.A--III.C,
respectively. Models 1 and 2 only allow for single quanta transitions (solid
arrows) while model 3 allows both single and double quanta transitions (dashed
arrows). (b) The double-path system for model 4 in Section 3.D.\bigskip}%
\end{center}
\end{figure}
%EndExpansion%
%TCIMACRO{\FRAME{ftbpFU}{4.1174in}{3.0885in}{0pt}{\Qcb{Power spectra of the
%control fields for model 1 aiming at a low yield of $O_{T}=5.0\%$ (Table III).
%$\gamma$ indicates the strength of decoherence (Eqs.(\ref{Liou1}) and
%(\ref{Liou2})). The control field intensity generally decreases with the
%increasing decoherence strength reflecting cooperative effects.}}{}%
%{fig2.eps}{\special{ language "Scientific Word";  type "GRAPHIC";
%width 4.1174in;  height 3.0885in;  depth 0pt;  original-width 4.0686in;
%original-height 3.4104in;  cropleft "0";  croptop "1";  cropright "1.1204";
%cropbottom "0";  filename 'Fig2.eps';file-properties "XNPEU";}}}%
%BeginExpansion
\begin{figure}
[ptb]
\begin{center}
\includegraphics[
trim=0.000000in 0.000000in -0.489859in 0.000000in,
height=3.0885in,
width=4.1174in
]%
{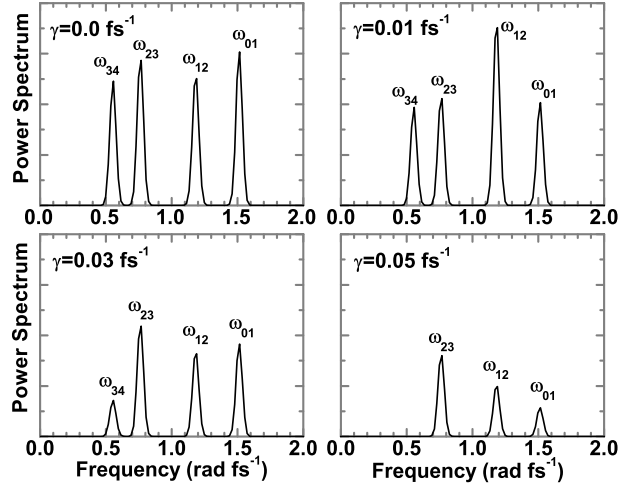}%
\caption{Power spectra of the control fields for model 1 aiming at a low yield
of $O_{T}=5.0\%$ (Table III). $\gamma$ indicates the strength of decoherence
(Eqs.(\ref{Liou1}) and (\ref{Liou2})). The control field intensity generally
decreases with the increasing decoherence strength reflecting cooperative
effects.}%
\end{center}
\end{figure}
%EndExpansion%
%TCIMACRO{\FRAME{ftbpFU}{4.1174in}{4.1174in}{0pt}{\Qcb{The same situation as
%Figure 2, but for model 2. Here the peaks corresponding to transitions
%involving higher quantum states disappear successively in keeping with the
%ability of decoherence to aid in meeting the desired physical goal.}}%
%{}{fig3.eps}{\special{ language "Scientific Word";  type "GRAPHIC";
%width 4.1174in;  height 4.1174in;  depth 0pt;  original-width 4.0562in;
%original-height 3.4104in;  cropleft "0";  croptop "1.1891";  cropright "1";
%cropbottom "0";  filename 'Fig3.eps';file-properties "XNPEU";}}}%
%BeginExpansion
\begin{figure}
[ptb]
\begin{center}
\includegraphics[
trim=0.000000in 0.000000in 0.000000in -0.644907in,
height=4.1174in,
width=4.1174in
]%
{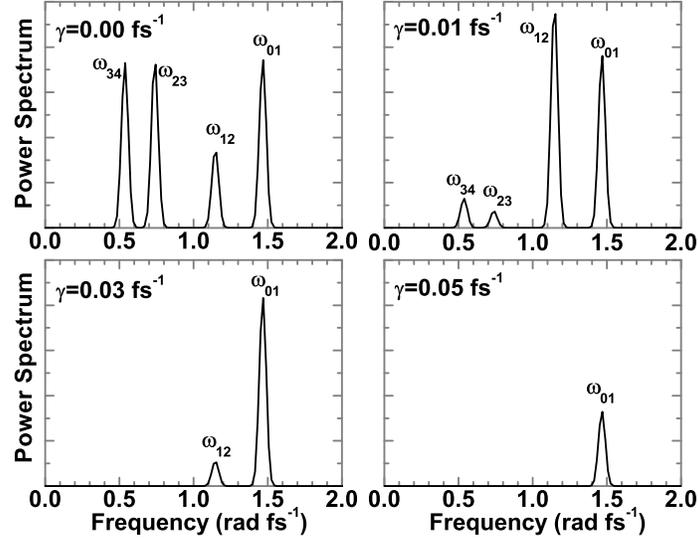}%
\caption{The same situation as Figure 2, but for model 2. Here the peaks
corresponding to transitions involving higher quantum states disappear
successively in keeping with the ability of decoherence to aid in meeting the
desired physical goal.}%
\end{center}
\end{figure}
%EndExpansion%
%TCIMACRO{\FRAME{ftbpFU}{4.1174in}{6.1769in}{0pt}{\Qcb{Power spectra of the
%control fields for model 4 (see panel (b) of Figure 1). The frequencies along
%the right path in Figure 1 are denoted with a prime to distinguish them from
%the left path. (a) No decoherence, (b) decoherence in the left path, (c)
%decoherence in the right path. In these low target yield cases the optimal
%field cooperates with the decoherence directing the dynamics to follow the
%associated path.}}{}{fig4.eps}{\special{ language "Scientific Word";
%type "GRAPHIC";  maintain-aspect-ratio TRUE;  display "USEDEF";
%valid_file "F";  width 4.1174in;  height 6.1769in;  depth 0pt;
%original-width 2.8587in;  original-height 3.4628in;  cropleft "0";
%croptop "1.2415";  cropright "1";  cropbottom "0";
%filename 'Fig4.eps';file-properties "XNPEU";}}}%
%BeginExpansion
\begin{figure}
[ptb]
\begin{center}
\includegraphics[
trim=0.000000in 0.000000in 0.000000in -0.836266in,
height=6.1769in,
width=4.1174in
]%
{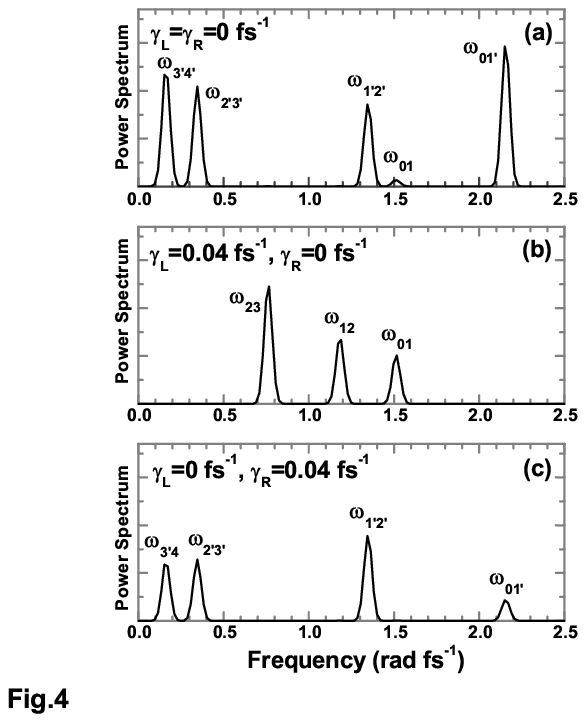}%
\caption{Power spectra of the control fields for model 4 (see panel (b) of
Figure 1). The frequencies along the right path in Figure 1 are denoted with a
prime to distinguish them from the left path. (a) No decoherence, (b)
decoherence in the left path, (c) decoherence in the right path. In these low
target yield cases the optimal field cooperates with the decoherence directing
the dynamics to follow the associated path.}%
\end{center}
\end{figure}
%EndExpansion

\end{document}